\documentclass[prl,twocolumn,showpacs,preprintnumbers,amsmath,amssymb]{revtex4}

\usepackage{graphicx}
\usepackage{amssymb,amsmath}
\usepackage{dcolumn}
\usepackage{bm}
\usepackage{color}

\begin{document}

\title{Optical analog of the Aharonov-Bohm interferometer}

\author{I.A. Shelykh}
\affiliation{Physics Department, University of Iceland, Dunhaga-3,
IS-107, Reykjavik, Iceland and St. Petersburg State Polytechnical
University, Polytechnicheskaya 29, 195251, St. Petersburg,
Russia}
\author{G. Pavlovic} \affiliation{LASMEA, UMR
CNRS-Universit\'e Blaise Pascal 6602, 24 Avenue des Landais, 63177
Aubi\`ere Cedex France}
\author{D. D. Solnyshkov} \affiliation{LASMEA, UMR
CNRS-Universit\'e Blaise Pascal 6602, 24 Avenue des Landais, 63177
Aubi\`ere Cedex France}
\author{G. Malpuech}
\affiliation{LASMEA, UMR CNRS-Universit\'e Blaise Pascal 6602, 24
Avenue des Landais, 63177 Aubi\`ere Cedex France}

\date{\today}

\begin{abstract}
We propose a novel spin-optronic device based on the interference of
polaritonic waves traveling in opposite directions and gaining
topological Berry phase. It is governed by the ratio of the TE-TM
and Zeeman splittings, which can be used to control the output
intensity. Due to the peculiar orientation of the TE-TM effective
magnetic field for polaritons, there is no analogue of the
Aharonov-Casher phaseshift existing for electrons.
\end{abstract}
\maketitle

The problem of the spin dynamics is one of the most interesting in
mesoscopic physics. The investigations in this field are stimulated
by the possibility of creation of nanodevices, where the spins of the
single particles could be precisely manipulated and controlled.
The first device of this type, namely spin transistor, was
proposed in early 90's in the pioneer work of Datta and Das
\cite{DattaDas}, who used an analogy between the precession of the
electron spin provided by Rashba spin-orbit interaction (SOI) and
the rotation of the polarization plane of light in optically anisotropic
media.

However, the experimental realization of the spin transistor
proposed by Datta and Das turned out to be quite complicated, due to
the extremely low efficiency of spin injection from ferromagnetic to
semiconductor materials. It was then proposed to use mesoscopic
gated Aharonov-Bohm (AB) rings as a possible basis of various
spintronic devices such as spin transistors \cite{Aronov,NittaAPL},
spin filters \cite{Frustaglia,KimKiselev,ShelykhSplitter}, and
quantum splitters \cite{ShelykhSplitter}. The conductance of such
structures depends both on the magnetic and the electric fields
applied perpendicular to the interface of the structure. The
magnetic field provides the AB phaseshift between the waves
propagating in the clockwise and anticlockwise directions thus
resulting in the oscillations of the conductance.

The electric field applied perpendicular to the plane of the ring
also affects the conductance. It has a two-fold effect. First, it
changes the carrier wavenumber thus leading to conductance
oscillations analogical to those observed in the Fabry-Perot
resonator. Second, it induces the Rashba
SOI and creates the dynamical phaseshift between the waves
propagating within the ring. It consists of Aharonov - Casher (AC)
phaseshift, arising from different values of the wavenumbers for the
waves propagating in opposite directions
\cite{NittaAPL,ShelykhRing}, and a Berry (geometric) phase
 term \cite{Aronov}, accumulated during the adiabatic evolution of the
electron's spin in the inhomogeneous effective magnetic field
created by Rashba SOI and external magnetic field perpendicular to
the structure's interface. As a result, the conductance of the
mesoscopic ring exhibits oscillations \cite{Molenkamp} as a function
of the perpendicular electric field.

It was recently proposed that in
the domain of mesoscopic optics
the controllable manipulation of the spin of excitons and
exciton-polaritons can provide a basis for the construction of
optoelectronic devices of the new generation, called spin-optronic
devices \cite{Spinoptronics}. The first element of this type,
polarization-controlled optical gate, was recently realized
experimentally \cite{OpticalGate}.

Exciton polaritons are the elementary excitations of
semiconductor microcavities in the strong coupling regime. Being
a mixture of quantum well (QW) excitons and cavity photons, they
possess a number of peculiar properties distinguishing them from
other quasi-particles in mesoscopic systems. An important property of
cavity polaritons is their (pseudo)spin \cite{ShelykhPSSb},
inherited from the spins of QW exciton and cavity photon and
directly connected with the polarization of emitted photons.
If one is able to control the spin of cavity polaritons, one can therefore
control the polarization of emitted light, which can be used in optical
information transfer.

The analog of Rashba SOI in microcavities is provided by the
longitudinal-transverse splitting (TE-TM splitting) of the polariton
mode. It is well known that due to the long-range exchange
interaction between the electron and the hole, for excitons having
non-zero in-plane wavevectors the states with dipole moment
oriented along and perpendicular to the wavevector are slightly
different in energy \cite{Maialle}. In microcavities, the splitting
of longitudinal and transverse polariton states is amplified due to
the exciton coupling with the cavity mode \cite{Panzarini1999} and
can reach values of about 1 meV.

TE-TM splitting results in the appearance of an effective magnetic
field provoking the rotation of polariton pseudospin. It is oriented
in the plane of the microcavity and makes a double angle with the $x$-axis in
the reciprocal space,

\begin{equation}
\vec{B}_{LT}(k)\sim\mathbf{e}_x\text{cos}(2\phi)+\mathbf{e}_y\text{sin}%
(2\phi) \label{EffectiveFieldTETM}
\end{equation}

This is different from the orientation of the effective magnetic
field provided by Rashba SOI (see Fig.1), which makes a single angle
with $y$-axis,

\begin{equation}
\vec{B}_{SOI}(k)\sim\mathbf{e}_x\text{sin}(\phi)-\mathbf{e}_y\text{cos}%
(\phi) \label{EffectiveFieldRashba}
\end{equation}

This peculiar orientation of  $\vec{B}_{LT}$ results in different
interference patterns for electrons and polaritons in ring
interferometers, leading in particular to the absence of AC
phaseshift for polaritons, as we shall see below \cite{Glazov}.

\begin{figure}[tbp]
\includegraphics[width=0.99\linewidth]{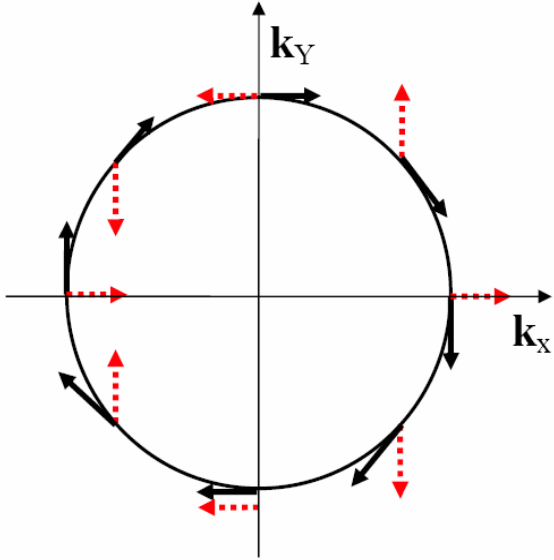}
\caption{ (color online) Orientation of the effective magnetic
fields provided by Rashba SOI (full arrows) and TE-TM (dashed arrows) splitting in the reciprocal
space} \label{EffFields}
\end{figure}

\begin{figure}[tbp]
\includegraphics[width=0.99\linewidth]{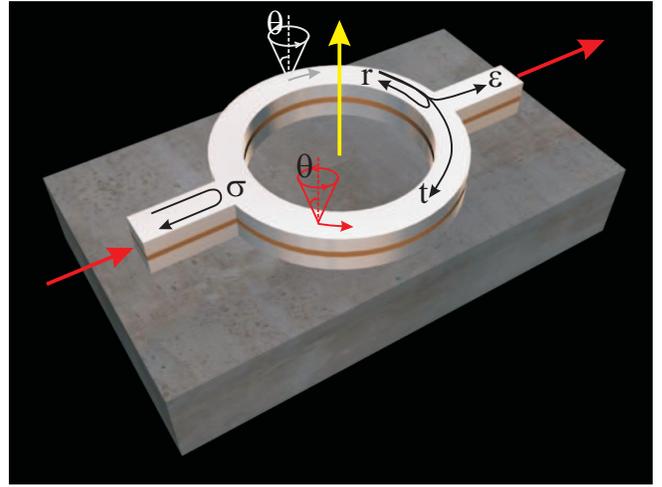}
\caption{ (Color online) Possible design of the microcavity waveguide. Letters mark the scattering amplitudes for particle propagation. Due to the flux conservation $\sigma^2+2\epsilon^2=1$ and $r^2+t^2+\epsilon^2=1$, see Ref.
\cite{Buttiker} for relevant details. Ellipses show the rotation of the pseudospin of polaritons
propagating along the arms of the ring. For clockwise and
anti-clockwise propagation the rotation direction of the
in-plane components of the pseudospin is different, which results in
different signs of corresponding Berry phases. Red arrows show the direction of light propagation, yellow arrow -- the direction of the magnetic field.
} \label{ABring}
\end{figure}

The system we consider in the present paper is an optical ring
interferometer placed in the external magnetic field perpendicular
to its interface (fig.~\ref{ABring}). The polaritons are injected in
the ingoing lead by a laser beam, propagate in the ring and leave it
by the outgoing lead, where their output intensity is detected. To make
the polaritons propagate along a desirable path, one needs to
engineer a corresponding confinement potential, which can be
achieved by variation of the cavity width
\cite{IdrissiKaitouni2006}, putting metallic stripes on the surface
of the cavity \cite{Lai2007} or applying a stress \cite{Balili2007}.
The other option is to produce the waveguide structure by
lithography, as in the case of micropillar cavities \cite{Bajoni}.
The narrow waveguide for polaritons has its own TE-TM splitting,
which is inversely proportional to waveguide dimensions
\cite{Dasbach} and which can dominate over
the cavity splitting (for a waveguide of $1 \mu$m width this
splitting can be as high as 1-2 meV \cite{Kuther}).

To calculate the intensity of the outgoing beam, we consider the
polariton states inside 1D ring, and take into account the TE-TM
splitting (of both origins) and the Zeeman splitting, the latter
provided by an external magnetic field perpendicular to the cavity
plane. With such geometry the averaging of the full Hamiltonian to 1D is valid since the energy splitting between two successive confined TE modes is orders of magnitude larger than the TE-TM splitting. The corresponding Hamiltonian in the basis of circular
polarized states reads \cite{Meijer}

\begin{eqnarray}
\widehat{H}=\left(\begin{array}{cc}
   H_0(\widehat{k})+\frac{\Delta_Z(B)}{2}& \frac{1}{2}[e^{-2i\phi},\Delta_{LT}(\widehat{k})]_+\\
  \frac{1}{2}[e^{2i\phi},\Delta_{LT}(\widehat{k})]_+ & H_0(\widehat{k})-\frac{\Delta_Z(B)}{2}\\
\end{array}\right)
\label{Hamiltonian}
\end{eqnarray}
where $\widehat{k}=-ia^{-1}d/d\phi$, $a$ is the radius of the ring,
$H_0(\widehat{k})$ is the bare polariton dispersion, $\Delta_{LT}(k)$ and $\Delta_{Z}(B)$ are the TE-TM and the Zeeman splittings respectively.
In our further consideration we use the effective mass approximation,
$H_0(\widehat{k})=\hbar^2\widehat{k}^2/2m_{eff}$ and assume the longitudinal-transverse splitting to be k-independent in the region of wavenumbers under study. The
solution of the Schroedinger equation with Hamiltonian
\eqref{Hamiltonian} can be expressed as (compare with a solution for
electrons \cite{Bulgakov})

\begin{eqnarray}
\Psi_+(\phi)=\frac{1}{\sqrt{1+\xi^2}}\left(\begin{array}{cc}-\xi
e^{+i\phi}\\e^{-i\phi}\end{array}\right)e^{ik_+a\phi} \label{Psi1}
\end{eqnarray}

\begin{eqnarray}
\Psi_-(\phi)=\frac{1}{\sqrt{1+\xi^2}}\left(\begin{array}{cc}
e^{+i\phi}\\ \xi e^{-i\phi}\end{array}\right)e^{ik_-a\phi}
\label{Psi2}
\end{eqnarray}
where the normalization factor reads

\begin{equation}
\xi=\frac{\Delta_{LT}/2\Delta_{Z}}{1+\sqrt{(\Delta_{LT}/2\Delta_{Z})^2+1}}
\label{normalization}
\end{equation}
and wavenumbers $k_\pm$ can be straightforwardly found from the
characteristic equation of the Hamiltonian \eqref{Hamiltonian}.

The eigenstates of the Hamiltonian \eqref{Hamiltonian} are elliptically
polarized. Their
pseudospin makes an angle
$\theta=\mathrm{arctan}(\Delta_{LT}/\Delta_Z)$ wit \emph{z}-azis. In the limit of weak
Zeeman splitting the polarization is linear ($\theta=0$), and in the
opposite limit it changes to circular ($\theta=\pi/2$).

The outgoing intensity can be found by
decomposing the states of ingoing and outgoing beams $\widehat{\Psi}_{in, out}$ by eigenstates of the Hamiltonian
\eqref{Hamiltonian} in the entrance and exit points (i.e. for
$\phi=0$ and $\phi=\pi$ respectively):

\begin{eqnarray}
\Psi_{in,out}=\frac{1}{\sqrt{1+|\xi|^2}}\left[A_{in,out}^+\left(\begin{array}{cc}\xi\\1\end{array}\right)+
A_{in,out}^-\left(\begin{array}{cc}-1\\\xi\end{array}\right)\right]
\label{PsiInOut}
\end{eqnarray}
where the outgoing amplitudes $A_{out}^{\pm}$ can be found as a sum of all
terms corresponding to the propagation between the ingoing and
outgoing leads. For a given spin orientation, the waves traveling in
the clockwise and anticlockwise directions obtain a different Berry
phase. Indeed, the direction of the effective magnetic field,
consisting of the in-plane TE-TM field and $z$-directed real field,
changes along the polariton trajectory and follows a cone-shaped
path (see Fig. \ref{ABring}). In the adiabatic approximation the
polariton pseudospin follows the direction of this field and the
corresponding geometric phase can be found as a half of the solid
angle covered by it \cite{Aronov},

\begin{equation}
\theta_B=\pm\pi\left(1-\frac{\Delta_Z}{\sqrt{\Delta_Z^2+\Delta_{LT}^2}}\right)
\end{equation}
where the sign corresponds to the propagation direction. One sees that $\theta_B$ depends on
the TE-TM and Zeeman splittings and changes from zero for
$\Delta_{LT}\ll\Delta_Z$ to $\pi$ for $\Delta_{LT}\gg\Delta_Z$. It differs by a factor of 2 (coming from Eq.\eqref{EffectiveFieldTETM})
from the geometric phase for electrons in the gated AB ring with Rashba SOI.

Considering only the processes with no more than one
round trip inside the ring, one has for outgoing amplitudes

\begin{eqnarray}
\nonumber \frac{A_{out}^{\pm}}{A_{in}^{\pm}\epsilon^2e^{-T/2\tau}e^{i\pi k_\pm
a}}&=&\left(1+r^2e^{-T/\tau}e^{2i\pi k_\pm
a}\right)cos(2\theta_B)+\\
&&+t^2e^{-T/\tau}e^{2i\pi k_\pm a}cos(6\theta_B)\label{AOut}
\end{eqnarray}

where $\tau$ is the polariton lifetime, $T=\pi a
\sqrt{m_{eff}}/\sqrt{2E}$ is the propagation time from ingoing to
outgoing lead, $\epsilon$ is the probability amplitude for a
polariton traveling along one of the arms of the ring to quit the
ring through the outgoing lead, $r$ is the amplitude of reflection
into the same arm, and $t$ is the amplitude of
transition into the other arm (see fig.~\ref{ABring}).

\begin{figure}[tbp]
\includegraphics[width=0.99\linewidth]{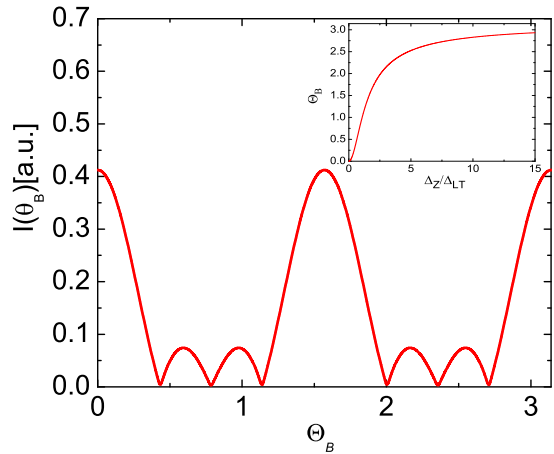}
\caption{(color online) The intensity of the outgoing polariton beam
as a function of the Berry phase $\theta_B$. The parameters are:
$T/\tau=0.05$, $r=0.15$ and $t=0.9$. The inset shows the dependence
of $\theta_B$ on $\Delta_Z/\Delta_{LT}$} \label{fig4}
\end{figure}

The formulae \eqref{PsiInOut} and \eqref{AOut} allow to determine
the intensity of the outgoing beam. It depends on the Berry phase
$\theta_B$, which thus plays a role of the AB phase of electronic
ring interferometers. The difference of the device we propose 
from the classical electronic AB interferometer is that
it needs the presence of the magnetic field and not just of the vector
potential in the region of the particle propagation.
It should also be noted that due to the peculiar
orientation of the TE-TM splitting for polaritons, there is no
analogue of the AC phaseshift present for electrons in a ring with
Rashba SOI. Indeed, for the electrons the AC phaseshift arises due
to the distinct wavenumbers for the particles traveling clockwise
and anticlockwise inside the ring, as for them the mutual
orientation of the spin and effective magnetic field provided by
Rashba SOI is different. On the contrary, for polaritons the
inversion of the propagation direction does not change the direction
of the effective magnetic field provided by TE-TM splitting (see
Fig. \ref{EffFields}), and thus the AC phaseshift is absent.

\begin{figure}[tbp]
\includegraphics[width=0.99\linewidth]{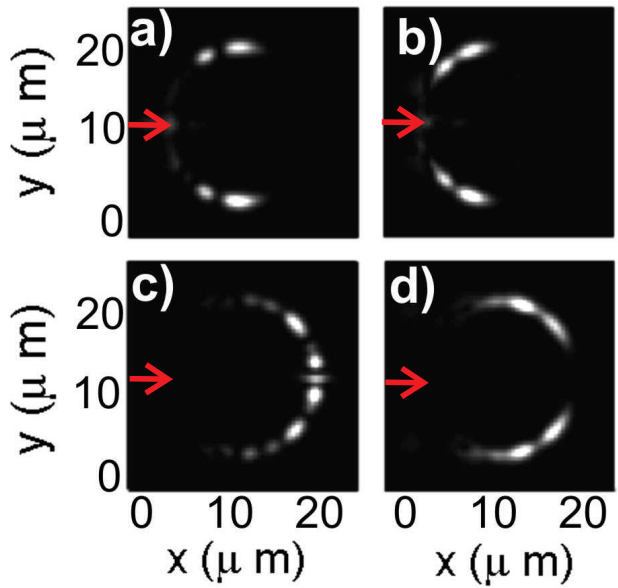}
\caption{Real-space images showing calculated emission intensity at Zeeman splitting 0~meV (a,c) and 2~meV (b,d) at different moments of time after the injection t=15~ps (a,b) and t=30~ps (c,d). Arrows show the direction of injection.}
\label{gp}
\end{figure}

The Berry phase can be modulated by tuning the intensity of TE-TM
splitting, e.g. by variation of the detuning between the exciton and
photon modes inside the ring (which can be achieved e.g. by
variation of stress forming the ring interferometer) or by tuning the
external magnetic field. The dependence of the intensity of the beam
on $\theta_B$ is shown at Fig. \ref{fig4}. One sees that the
intensity is maximal for $\theta_B=0,\pi$ and in the region between
these two points reveals a local maximum at $\theta_B=\pi/2$.

To check the results of our analytical theory, we have performed a
simulation of the structure we propose using coupled
Gross-Pitaevskii equations for excitons and Schroedinger equations
for photons taking into account their polarization
\cite{ShelykhPRL}. In this simulation we studied pulse propagation
through the ring interferometer of $16~\mu$m diameter without
magnetic field as well as under a field of 35 Tesla causing an
exciton Zeeman splitting of 2 meV. This value can be sufficiently
reduced by the use of diluted semi-magnetic cavities
\cite{Brunetti} or by choosing a different material system (e.g. CdSe/ZnSe \cite{Renner}).

The results of our simulations are shown in Figure~\ref{gp}. All
images show spatial distribution of the emission intensity, which
is directly proportional to the local density of polaritons. The top
two panels (a,b) show the initial stage of the pulse propagation
through the ring waveguide at time $t$=15 ps after the excitation.
The bottom two panels (c,d) show the final stage of the pulse
propagation ($t$=30~ps), when the two beams interfere at the outgoing
lead connection point. Without magnetic field (left column -- panels
a,c) the interference is constructive, and the output into the
outgoing lead has the highest value. Under a certain magnetic field
(right column -- panels b,d) the interference is destructive and a
dark spot is visible instead of a bright one. This result
corresponds to the predictions of the analytical theory and
demonstrates that such a waveguide can indeed operate as an optical
interferometer.

In conclusion, we have proposed an optical analog of the
spin-interference device based on a mesoscopic ring interferometer.
We demonstrated that the Berry phase provided by the TE-TM and Zeeman splittings
for polaritons  plays a role of AB phase for electrons
and leads to a variation of the intensity of the
outgoing beam. On the other hand there is no analogue of the corresponding Aharonov-Casher effect because of the peculiar symmetry of the TE-TM splitting. This system allows to solve the main difficulties occurring in electronic systems such as the low efficiency of spin injection from a ferromagnet to a semiconductor system. The effect we propose cannot be observed for bare cavity photons, but is specific of strongly coupled exciton-polaritons because it requires a finite Zeeman splitting. The use of exciton-polaritons is also highly advantageous with respect to the bare excitons \cite{Nio}, since the mean free path of exciton-polaritons is much longer due to their photon component.

The authors thank M. Glazov, N. Gippius, and S. Tikhodeev for fruitful discussions and critical reading of the manuscript.
The work was supported by the ANR Chair of Excellence Program and by
the EU STIMSCAT FP6-517769 project. I.A. Shelykh acknowledges the
support from the Grant of the President of Russian Federation.


\begin{thebibliography}{99}

\bibitem{DattaDas} S. Datta and B. Das, \emph{Appl. Phys. Lett.} \textbf{56}, 665 (1990).

\bibitem{Aronov} A.G. Aronov and Y.B. Lyanda-Geller. \emph{Phys. Rev. Lett.} \textbf{70}, 343 (1993).

\bibitem{NittaAPL} J. Nitta \emph{et al}, \emph{Appl. Phys. Lett.} \textbf{75}, 695 (1999).

\bibitem{Frustaglia} M. Popp \emph{et al} \emph{Nanotechnology} \textbf{14}, 347
(2003).

\bibitem{KimKiselev} A.A. Kiselev, K.W. Kim, \emph{J. Appl. Phys.} \textbf{94}, 4001 (2003).

\bibitem{ShelykhSplitter} I.A. Shelykh \emph{et al}, \emph{Phys. Rev. B} \textbf{72}, 235316 (2005).

\bibitem{ShelykhRing} I.A. Shelykh et al, \emph{Phys. Rev. B} \textbf{71}, 113311 (2005).

\bibitem{Molenkamp} M. Konig \emph{et al}, \emph{Phys. Rev.
Lett.} \textbf{96}, 076804 (2006).

\bibitem{Spinoptronics} I. Shelykh \emph{et al}, \emph{Phys. Rev. B} \textbf{70}, 035320 (2004).

\bibitem{OpticalGate} C. Leyder \emph{et al}, \emph{Phys. Rev. Lett.} \textbf{99}, 196402 (2007).

\bibitem{ShelykhPSSb} I.A. Shelykh \emph{et al}, \emph{Phys. Stat. Sol. (b)}
\textbf{242}, 2271 (2005).

\bibitem{Maialle} M. Z. Maialle \emph{et al}, \emph{Phys. Rev. B} \textbf{47}, 15776 (1993).

\bibitem{Panzarini1999} G. Panzarini \emph{et al}, \emph{Phys. Rev. B }\textbf{59},
5082 (1999).

\bibitem{Glazov} One of the consequences of the absence of the AC phase for polaritons is a lack of the weak antilocalization for them, see M. M. Glazov and L. E. Golub,
\emph{Phys. Rev. B} \textbf{77}, 165341 (2008).

\bibitem{IdrissiKaitouni2006}
R. Idrissi Kaitouni, {\it et. al.}, Phys. Rev. B, {\bf 74}, 155311
(2006).

\bibitem{Lai2007}
C. W. Lai, {\it et. al.}, Nature, {\bf 450}, 529 (2007).

\bibitem{Balili2007} R. Balili et al, \emph{Science} \textbf{316}, 1007 (2007).
\bibitem{Bajoni} D. Bajoni et al, Phys. Rev. Lett. \textbf{100}, 047401 (2008).
\bibitem{Dasbach} G. Dasbach et al, Phys. Rev. B \textbf{71}, 161308R (2005).
\bibitem{Kuther} A. Kuther et al, Phys. Rev. B \textbf{58}, 15744 (1998).

\bibitem {Meijer} Compare with Hamiltonian for electrons, F. E.
Meijer \emph{et al}, \emph{Phys. Rev. B} \textbf{66}, 033107 (2002)

\bibitem{Bulgakov} E.N. Bulgakov, A.F. Sadreev, \emph{Phys. Rev. B} \textbf{66}, 075331 (2002).

\bibitem{Buttiker} M. Buttiker \emph{et al}, \emph{Phys. Rev. A} \textbf{30}, 1982 (1984).

\bibitem{ShelykhPRL} I.A. Shelykh \emph{et al}, \emph{Phys. Rev. Lett.} \textbf{97}, 066402 (2006).

\bibitem{Brunetti} A. Brunetti \emph{et al}, \emph{Phys. Rev. B} \textbf{73}, 205337 (2006)

\bibitem{Renner} J. Renner \emph{et al}, \emph{Appl. Phys. Lett.} \textbf{93}, 151109 (2008).

\bibitem{Nio} Wan Nio and Qian Niu, \emph{Phys Rev. Lett.} \textbf{101}, 106401 (2008).


\end{thebibliography}
\end{document}